\RequirePackage{fix-cm}

\documentclass[twocolumn]{svjour3}          
\smartqed  
\usepackage{amsmath}
\usepackage{amssymb}
\usepackage{graphicx}
\usepackage{marvosym}
\usepackage{mathtools}
\usepackage{xfrac}
\usepackage{bbm}
\usepackage{natbib}

\usepackage{tikz}
\usepackage{pgfplots}
\pgfplotsset{compat=newest}
\usepgfplotslibrary{groupplots}
\usepgfplotslibrary{external} 
\newlength\figurewidth
\newlength\figureheight
\definecolor{cl1}{RGB}{189,189,189}
\definecolor{cl2}{RGB}{99,99,99}
\definecolor{cl3}{RGB}{0,0,0}

\newcommand{\df}{\vcentcolon =}

\newcommand{\N}{{\mathbb N}}

\newcommand{\EV}[1]{\mathbb{E}{\kern -0.15em}\left[ #1 \right]}
\newcommand{\cEV}[2]{\mathbb{E}{\kern -0.15em}\left[ #1\, \middle| \, #2 \right]}
\newcommand{\Prob}[1]{\mathbb{P}{\kern -0.15em}\left[ #1 \right]}
\newcommand{\cProb}[2]{\mathbb{P}{\kern -0.15em}\left[ #1 \, \middle| \, #2 \right]}

\newcommand{\Var}[1]{\mathrm{Var}{\kern -0.15em}\left[ #1 \right]}

\newcommand{\abs}[1]{\lvert #1 \rvert}
\newcommand{\norm}[1]{\lVert #1 \rVert}

\newcommand{\scp}[2]{\langle #1, #2 \rangle}

\renewcommand{\phi}{\varphi}

\newcommand{\dt}{\,\mathrm{d}t}
\newcommand{\ds}{\,\mathrm{d}s}
\newcommand{\dr}{\,\mathrm{d}r}

\newcommand{\dx}{\,\mathrm{d}x}

\newcommand{\dwt}{\,\mathrm{d}W(t)}

\newcommand{\dxt}{d_{\hat{X}}(t)}

\begin{document}

\title{Reliability of signal transmission in stochastic nerve axon equations\thanks{This work is supported by the BMBF, FKZ 01GQ1001B.}}
\author{Martin Sauer \and Wilhelm Stannat}

\institute{M. Sauer, W. Stannat (\Letter) \at%
Institut f\"ur Mathematik, Technische Universit\"at Berlin,\\
Stra\ss e des 17. Juni 136, D-10623 Berlin, Germany and\\
Bernstein Center for Computational Neuroscience,\\
Philippstr. 13, D-10115 Berlin, Germany.\\
\email{stannat@math.tu-berlin.de}}

\date{December 2, 2015}

\journalname{Preprint}
\maketitle

\begin{abstract}
We introduce a method for computing probabilities for spontaneous activity and propagation 
failure of the action potential in spatially extended, conduc\-tance-based neuronal models 
subject to noise, based on statistical properties of the membrane potential. We compare 
different estimators with respect to the quality of detection, computational costs and 
robustness and propose the integral of the membrane potential along the axon as an 
appropriate estimator to detect both spontaneous activity and propagation failure. Performing a model reduction we achieve a simplified analytical expression based on the linearization at the resting potential (resp. the traveling action potential). This allows to approximate the probabilities for spontaneous activity and propagation failure in terms of (classical) hitting probabilities of one-dimensional linear stochastic differential equations. The quality of the approximation with respect to the noise amplitude is discussed and illustrated with numerical results for the spatially extended Hodgkin-Huxley equations. Python simulation code 
is supplied on GitHub under the link https:// github.com/deristnochda/Hodgkin-Huxley-SPDE. 

\end{abstract}

\section{Introduction}
Noise is an inherent component of neural systems that accounts for various problems in information processing at all levels of the nervous system, see e.\,g. the review \cite{FaisalNature} for a detailed discussion. In particular, channel noise has been identified as an important source of various types of variability in single neurons. Examples are the noise induced phenomena as observed in \cite{Faisal}. The timing of action potentials can be highly sensitive with respect to fluctuations in the opening and closing of ion channels leading to jitter and stochastic interspike intervals \citep{Horikawa}. This effect becomes important in thin axons with diameter of less than $1\mu$m. Furthermore, there appear stochastic patterns in the grouping of action potentials, and action potentials can vanish due to noise interference or spontaneously emerge without apparent synaptic input.

When it comes to the mathematical modeling of the membrane potential in axons, in particular in thin ones, channel noise therefore has to be taken into account. For a discussion and comparison of the various types of adding noise to conductance-based neuronal models such as the classical Hodgkin-Huxley equations we refer to \cite{GoldwynSheaBrown}. Concerning spatially extended models, in e.\,g. \cite{TuckwellJost, TuckwellJost2, TuckwellFHN} it has been shown that already simple additive noise, uncorrelated in space and time, accounts for a large range of variability in the action potential. That includes variability in the repetitive generation of action potentials, deletion of action potentials or {\em propagation failures} and spontaneously emerging action potentials or {\em spontaneous activity}. 
Because of this observation, we restrict ourselves to such a simple model of the noise that — as a byproduct — reduces the computational and analytical complexity. However, the proposed detection and estimation method can be applied to more complex models and e.g. even the full Markov chain dynamics of channel noise can be used.

It is the purpose of this work to introduce a method to compute in a mathematical consistent way the probabilities of those last two events. This is done for general spatially extended neuronal models with additive noise, both numerically and theoretically, in terms of statistical quantities of the membrane potential. A suitable statistical estimator for such kind of characteristics should have the following desired properties: It is automatically evaluable to do Monte-Carlo simulations; it strictly separates the considered event from different ones; it is a low dimensional function of the observables; it is relatively robust to stochastic perturbations and uncertainty in the observables. We compare different estimators with respect to the quality of detection, computational costs and robustness. In order to further reduce the computational costs and to obtain a simpler analytical description, we perform a consistent model reduction, with respect to these statistical quantities, to a one-dimensional linear stochastic differential equation that allows to compute the desired characteristics without necessarily simulating the full system.

The method is illustrated in a case study using the Hodgkin-Huxley equations \citep{HodgkinHuxley} with two distinct parameter sets. With spatial diffusion, this is a system of partial differential equations that can serve as a model for the propagation of action potentials in the neuron's axon. In particular, depending on the size of the stimulus there exist pulse-like solutions (action potentials) to these equations propagating along the spatial domain. Using these equations, we estimate the probabilities of {\em spontaneous activity} and {\em propagation failure}. Although we only focus on these two examples, the methods presented here can be used for a broader range of problems, in particular, similar model reductions can also be performed in order to compute time jitter and the variability in grouping patterns of action potentials.

In our setting, we consider a simple spatial geometry of the axon that is a cylindrical shaped fiber. Thus the relevant spatial domain is an interval $[0,L]$. We propose $\Phi (u) \df \int_0^L u(x) \dx$ as an estimator for the detection of spontaneous activity and propagation failures. Here, $u$ is the space(-time)-dependent observable whose solution is pulse-formed. In the cases at hand, this will be the membrane potential. $\Phi(u)$ is the area under the pulse considered as a graph with respect to the space variable that has the following properties: It is easy to extract automatically from the numerical simulations; it significantly separates the number of observed pulses; it is a linear functional of only one observable;  stochastic perturbations, in particular additive noise that is white in space (or of low correlation length) should cancel out through integration. The events of {\em spontaneous activity} and {\em propagation failure} can both be defined as threshold crossings of the quantity $\Phi(u)$ and therefore easily be estimated using a Monte-Carlo simulation. The results can be found in Section 3. In Section 4, we do a model reduction for this quantity, only assuming a reasonable local stability of the pulse and resting solutions. In particular, we deduce one-dimensional Ornstein-Uhlenbeck processes, that captures both probabilities in particular for small noise intensities remarkably well. 

\section{Hodgkin-Huxley type equations}
In this article, we consider a spatially extended conductance based neuronal model with a simple one dimensional domain $(0,L)$ approximating the axon. This is most accurate in the case of a long axon, shaped as a cylinder with constant diameter. Our examples combine a Hodgkin-Huxley type model with diffusive spatial coupling to describe the evolution of the membrane potential $u(t,x)$ in time and space by a system of partial differential equations involving the dimensionless potassium activation, sodium activation and sodium inactivation variables $n(t,x)$, $m(t,x)$ and $h(t,x)$, respectively. This typically reads as
\begin{equation}\label{eq:pdeHoHu}
\begin{aligned}
C_m \partial_t u &= \tfrac{d}{4R_i} \partial_x^2 u - \overline{g}_{\text{K}} n^4 (u - E_{\text{K}})\\
&\quad - \overline{g}_{\text{Na}} m^3 h ( u - E_{\text{Na}}) - g_{\text{L}} (u - E_{\text{L}})\\
\partial_t n &= \alpha_n (u) ( 1 - n) - \beta_n(u) n,\\
\partial_t m &= \alpha_m (u) ( 1 - m) - \beta_m (u) m,\\
\partial_t h &= \alpha_h (u) ( 1 - h) - \beta_h(u) h.
\end{aligned}
\end{equation}
Here, $C_m$ is the membrane capacitance in $\mu\text{F}/\text{cm}^2$, $d$ the axon diameter in cm, $R_i$ the intracellular resistivity in $\Omega$cm, $g_{\text{K}}, g_{\text{Na}}, g_{\text{L}}$ the maximal potassium, sodium and leak conductance in mS$/\text{cm}^2$. To further specify units, all times are in ms, voltages in mV and distances in cm. These standard parameters from the original work of \cite{HodgkinHuxley} are used throughout: $R_i = 34.5$, $C_m = 1$, $g_{\text{K}} = 36$, $g_{\text{Na}} = 120$, $g_{\text{L}} = 0.3$, $E_{\text{K}} = -12$, $E_{\text{Na}} = 115$ and $E_{\text{L}} = 10$. Note that the membrane potential is shifted by $65$mV compared to the original values. In order to be in the regime of thin, unmyelinated axons, we choose a diameter of $d=0.5 \mu$m for all simulations and consider an axon length of $L=1$cm.

\subsection{Two parameter sets for the (in)activation variables}

Equation \eqref{eq:pdeHoHu} is missing the coefficients determining the evolution of the (in)activation variables. In the standard model following \cite{HodgkinHuxley} these are
\begin{align*}
\alpha_n(u) &= \frac{10-u}{100\big( \mathrm{e}^{\frac{10-u}{10}} - 1 \big)}, &\beta_n(u) &= \frac{1}{8} \mathrm{e}^{-\frac{u}{80}},\\
\alpha_m(u) &= \frac{25-u}{10\big( \mathrm{e}^{\frac{25-u}{10}} - 1 \big)}, &\beta_m(u) &= 4 \mathrm{e}^{-\frac{u}{18}},\\
\alpha_h(u) &= \frac{7}{100} \mathrm{e}^{-\frac{u}{20}}, &\beta_h(u) &= \frac{1}{\mathrm{e}^{\frac{30-u}{10}} + 1}. 
\end{align*}
In the following, we refer to this model as (HH). A second model ($\widetilde{\mathrm{HH}}$) with a different behavior can be obtained by slight modification. Set
\[
\tilde{\alpha}_m(u) = \frac{36-u}{10\big( \mathrm{e}^{\frac{36-u}{10}} - 1 \big)}, \quad \tilde{\beta}_h(u) = \frac{1}{\mathrm{e}^{\frac{21.5-u}{10}} + 1},
\]
that amounts to a change in the sensitivity of the sodium (in)activation rates, and leave the rest unchanged. The result is a neuron much less sensitive to input, i.\,e. with a higher firing threshold. In the next section, models (HH) and ($\widetilde{\text{HH}}$) are used to illustrate the phenomenon of {\em spontaneous activity} and {\em propagation failure}, respectively.

\subsection{A mathematical model}

Noisy perturbations of equation \eqref{eq:pdeHoHu} can be realized as a stochastic partial differential equation (SPDE) on the Hilbert space $(H, \norm{\cdot}) = L^2(0,L)$ with inner product $\scp{\cdot}{\cdot}$. The variables $u(t), n(t), m(t)$ and $h(t)$ are then function valued, thus we omit the $x$ dependence in the notation.

For the spatial diffusion, define  the Laplace operator $\Delta u \df \partial^2_{x} u$ supplemented with Neumann boundary conditions. We choose a sealed end at $x = L$, i.\,e. $\partial_x u(t,L) = 0$ for all $t \geq 0$ and model the input signal to the axon via an injected current in form of a rectangular pulse
\begin{equation}\label{eq:DefInput}
\partial_x u(t, 0) = -\frac{4R_i J(t)}{\pi d^2}, \quad J(t) = \begin{cases} J, & t \leq T^\ast\\ 0, & t > T^\ast \end{cases}
\end{equation}
Here, $T^\ast \leq \infty$ is the duration and $J > 0$ the amplitude of the signal.

The question of how to add noise to equation \eqref{eq:pdeHoHu} has been studied in the literature, see e.\,g. \cite{GoldwynSheaBrown}. Although it has been shown that current noise, i.\,e. uncorrelated additive noise in the voltage variable, does not accurately approximate a Markov chain ion channel dynamics, we use this form of noise in our study. The reason is twofold: First, already such a kind of noise can qualitatively account for all of the phenomena observed in e.\,g. \cite{Faisal} and second, it allows further analysis due to its simplicity. Mathematically speaking, current noise is realized as a two-parameter white noise $\eta$ that is defined in terms of a cylindrical Wiener process $W$ such that $\eta = \dot{W}$. $W = (W(t))_{t\geq 0}$ is a function valued process that can be formally represented by the infinite series
\[
W(t)(x) \df \sum_{n=1}^\infty e_n(x) \beta_n(t),
\]
where $(\beta_n(t))_{n\in \N}$ is a family of iid real valued Brownian motions and
\[
e_n(x) \df \sqrt{\frac{2}{L}} \cos \Big(2\pi \frac{n}{L} x \Big)
\]
is an orthonormal basis of $H$. For $f,g \in H$ one can calculate the covariance of this process as
\[
\EV{ \scp{W(t)}{f} \scp{W(s)}{g}} = (t \wedge s) \scp{f}{g},
\]
thus $\EV{\eta(t,x) \eta(s,y)} = \delta(t-s) \delta(x-y)$, i.\,e. no correlation in either time nor space. Thus formally speaking, a cylindrical Wiener process is time-integrated space-time white noise. Equation \eqref{eq:pdeHoHu} then reads as
\begin{equation}\label{eq:spdeHoHu}
\begin{aligned}
\mathrm{d}u(t) &= \big[\lambda \Delta u(t) +f \big( u(t), n(t), m(t), h(t)\big)\big] \dt\\
&\qquad + \sigma \dwt,\\
\tfrac{\mathrm{d}n(t)}{\dt} &= \alpha_n \big(u(t)\big) \big( 1 - n(t)\big) - \beta_n\big(u(t)\big) n(t),\\
\tfrac{\mathrm{d}m(t)}{\dt} &= \alpha_m \big(u(t)\big) \big( 1 - m(t)\big) - \beta_m\big(u(t)\big) m(t),\\
\tfrac{\mathrm{d}h(t)}{\dt} &= \alpha_h \big(u(t)\big) \big( 1 - h(t)\big) - \beta_h\big(u(t)\big) h(t).
\end{aligned}
\end{equation}

Together with suitable initial conditions, in our case the equilibrium values $(u^\ast, n^\ast, m^\ast, h^\ast)$, being $u^\ast = 0$ for (HH) and $u^\ast \approx -0.820$ for ($\widetilde{\text{HH}}$), as well as
\[
x^\ast = \frac{\alpha_x(u^\ast)}{\alpha_x(u^\ast) + \beta_x(u^\ast)}, \quad x=n,m,h,
\]
we refer to \cite{SauerNerveAxon} for well-posedness of equation \eqref{eq:spdeHoHu}.

\subsection{Linear stability of pulse and resting state}
If one injects an input above a certain threshold, the solution of equation \eqref{eq:pdeHoHu} rapidly approaches a traveling pulse like solution. Denote $X=(u,n,m,h)^T$, then numerical simulations show that  this traveling pulse is well-approximated by a solution of the form 
$X(t,x) = \hat{X}(x - ct)$ for a fixed reference profile $\hat{X}$ and pulse speed $c$ as long as the pulse did not reach the boundary. Let 
us call this solution $\hat{X}(t)$.

Without any external input, the system \eqref{eq:pdeHoHu} remains in equilibrium if started there. Denote by $X^\ast$ this constant (in time and space) solution to the equations.

The phenomena of interest in this work directly correspond to the stability properties of those two solutions $\hat{X}$ and $X^\ast$. Although this has only been shown for general stochastic bistable equations, see e.\,g. \cite{StannatBistable}, we assume a linear stability condition that should be possible to be extended to the higher dimensional Hodgkin-Huxley system. This linear stability assumption is then only used in Section 4 for a model reduction. For convenience of notation, denote equation \eqref{eq:spdeHoHu} in the following abstract form
\begin{equation}\label{eq:AbstractSPDE}
\mathrm{d} X(t) = \Big( A X(t) + F\big(X(t)\big)\Big)\dt + \sigma \,\mathrm{d}\mathbb{W}(t),
\end{equation}
where $A = (\Delta, 0, \dots)^T$, $\mathbb{W} = (W, 0, \dots)^T$ and $F$ is the appropriate nonlinear part of the drift. Also, denote by $\mathcal{H} = \otimes_{n=1}^4 H$ the state space of \eqref{eq:AbstractSPDE}. Then, we assume the following geometrical condition of Lyapunov type
\begin{equation}\label{eq:StabilityRest}
\scp{\big[ A + \nabla F\big(X^\ast\big)\big] h}{h}_\mathcal{H} \leq -\kappa^\ast \norm{h}_\mathcal{H}^2,
\end{equation}
implying that the resting solution is locally exponentially attracting in $\mathcal{H}$, i.\,e. linearly stable. Moreover
\begin{equation}\label{eq:StabilityPulse}
\scp{\big[ A + \nabla F\big(\hat{X}(t)\big)\big] h}{h}_\mathcal{H} \leq -\hat{\kappa} \norm{h}_\mathcal{H}^2 + \hat{C} \scp{h}{\dxt}_\mathcal{H}^2,
\end{equation}
for all $t \in [T_0, T]$, where $\dxt = \dot{\hat{X}}(t)$. Here $T_0$ is the time until $\hat{X}$ is in pulse form and $T$ denotes the time, when the pulse has reached the boundary. The latter condition can be interpreted geometrically as follows: once it is formed, the traveling pulse solution is locally exponentially attracting in the subspace $\bot_t \df \{ h \in \mathcal{H}: \scp{h}{\dxt}_\mathcal{H} = 0\} \subset \mathcal{H}$ that is orthogonal to the direction of propagation.

\subsection{Numerical method}

SPDE \eqref{eq:spdeHoHu} is a reaction diffusion equation coupled to a set of equations without spatial diffusion. Thus, the main issue from a numerical perspective is the simulation of equations of the form
\[
\mathrm{d} u(t) = \big[ \lambda\Delta u(t) + f\big(t, u(t)\big) \big] \dt + \sigma \dwt
\]
with Neumann boundary conditions as in \eqref{eq:DefInput}. The numerical method chosen for the integration of such a SPDE is a finite difference approximation in both space and time, see \cite{SauerLattice,SauerNerveAxon} for details. For the space variable $x$ we use an equidistant grid $(x_i)$ of size $\Delta x = \sfrac{L}{N}$ and replace the second derivative by its two-sided difference quotient. Boundary conditions are approximated up to second order, using the artificial points $x_{-1}$ and $x_{N+1}$. The time variable $t$ is discretized to $(t_j)$ using $\Delta t = \sfrac{1}{M}$ and a semi-implicit Euler scheme. Approximating the variable $u$ in the point $(x_i, t_j)$ yields the following scheme.
\begin{align*}
u_{0,j+1} &= u_{0,j} + \tfrac{\lambda \Delta t}{\Delta x^2} \big( 2u_{1,j+1} - 2 u_{0,j+1}\big)\\
&\qquad + \Delta t f \big( u_{0,j} \big) + 2 \tfrac{\lambda \Delta t}{\Delta x} J_{j+1} + \sigma \sqrt{\tfrac{\Delta t}{2\Delta x}} N_{0,j},\\
u_{i,j+1} &= u_{i,j} + \tfrac{\lambda \Delta t}{\Delta x^2} \big( u_{i+1,j+1} - 2 u_{i,j+1} + u_{i-1,j+1} \big)\\
&\qquad + \Delta t f \big( u_{i,j} \big) + \sigma \sqrt{\tfrac{\Delta t}{\Delta x}} N_{i,j},\\
u_{N,j+1} &= u_{N,j} + \tfrac{\lambda \Delta t}{\Delta x^2} \big( 2u_{N-1,j+1} - 2 u_{N,j+1} \big)\\
&\qquad + \Delta t f \big( u_{N,j} \big) + \sigma \sqrt{\tfrac{\Delta t}{2\Delta x}} N_{N,j}
\end{align*}
for $1 \leq i \leq N-1$, where $J_{j}$ is the discrete applied current and $(N_{i,j})_{0\leq i\leq N, j\geq 1}$ is a sequence of iid $\mathcal{N}(0,1)$-distributed random variables. For details on convergence of this scheme and error rates we refer to \cite{SauerLattice}.

\section{Reliability of signal transmission}\label{sec:Reliability}

\begin{figure}[ht]
\centering
\includegraphics{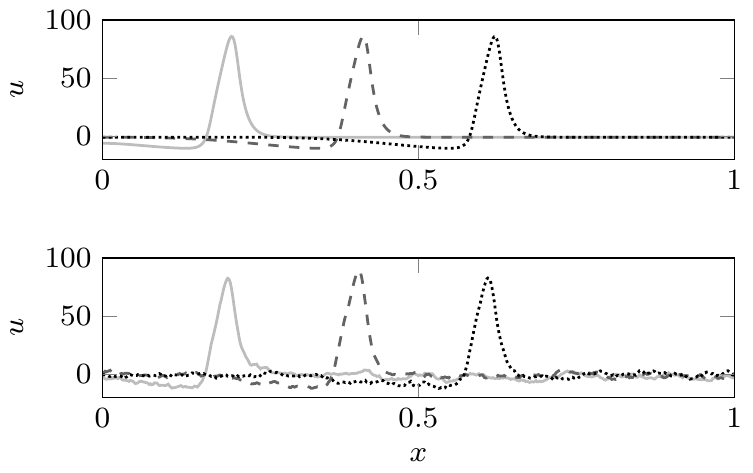}
\caption{\small The time evolution of $u$ using ($\widetilde{\text{HH}}$) at $t_1 = 10$ (solid), $t_2 = 20$ (dashed) and $t_3 = 30$ (dotted) for the deterministic pulse ($\sigma=0$) and one perturbed by noise ($\sigma=0.5$).}
\label{fig:EvolutionPulse}
\end{figure}

Let us first specify numerical parameters. We use $N=500$ gridpoints, i.\,e. $\Delta x = 0.02$, and $\Delta t = 0.01$ to simulate the equations. Using the input of height $J = 0.001 \mu$A and length $T^\ast = 0.5$, in both models (HH) and ($\widetilde{\text{HH}}$) a pulse is formed at the left boundary, traveling to the right, see Figure \ref{fig:EvolutionPulse}.

\begin{figure}[ht]
\centering
\includegraphics{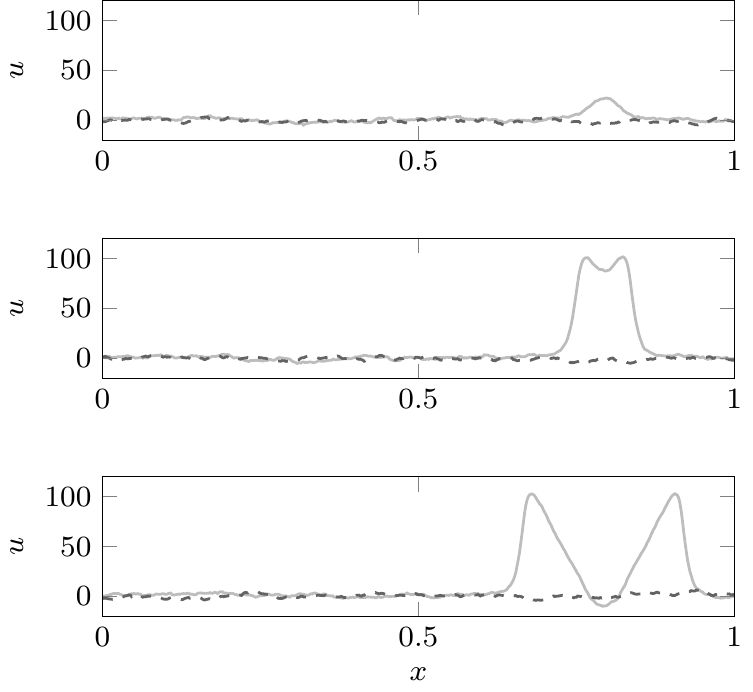}
\caption{\small A realization of the event {\em spontaneous activity} is given by the solid, light gray trajectory. The three plots are the membrane potential $u$ using (HH) at times $t_1 = 13.5$, $t_2 = 14.5$, $t_3=16.5$ from top to bottom. For comparison we include a trajectory, where there are only fluctuations around the resting potential (dashed, dark gray). For all of them, $\sigma=0.372$.}
\label{fig:sa_example}
\end{figure}

The problem at hand is how the presence of noise affects the generation and reliability of transmission of action potentials in the axon, similar to the studies by \cite{Faisal} for the Hodgkin-Huxley equations and \cite{TuckwellFHN} for the FitzHugh-Nagumo equations. In particular, this section concerns two distinct phenomena observed in these two studies. Faisal \& Laughlin found that in the (HH) model action potential propagation is very secure, but in certain cases there spontaneously emerge action potentials somewhere along the axon due to the effect of noise ({\em spontaneous activity}). This is illustrated in Figure \ref{fig:sa_example}, where an exemplary trajectory of such an event can be found.

\begin{figure}[ht]
\centering
\includegraphics{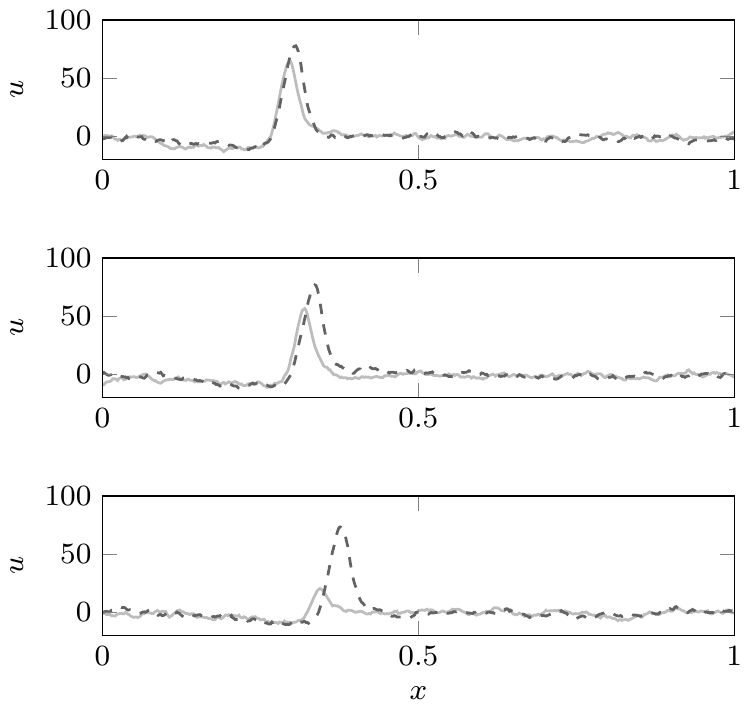}
\caption{\small A realization of the event {\em propagation failure} is given by the solid, light gray trajectory. The three plots are the membrane potential $u$ using ($\widetilde{\text{HH}}$) at times $t_1 = 14.5$, $t_2 = 16$, $t_3=18$ from top to bottom. For comparison we include a trajectory, where no propagation failure occurs (dashed, dark gray). For all of them, $\sigma=0.504$.}
\label{fig:pf_example}
\end{figure}

On the other hand, Tuckwell observed that a primary effect of noise on the action potential can be a breakdown of the pulse without any secondary phenomena such as spontaneous activity ({\em propagation failure}). An illustration is given in Figure \ref{fig:pf_example} comparing a failure to a stable pulse. The equations Tuckwell used to model the neuron are, of course, different to the work by Faisal \& Laughlin, however this discrepancy is not due to the choice of the neuron model but rather due to the choice of the particular parameter values describing the model. These are directly linked to the stability of the traveling pulse and resting state. Indeed, slightly modifying sodium (in)activation in model ($\widetilde{\text{HH}}$), we can observe occurences of propagation failure but no spontaneous activity. 
In this work, (HH) is always used to study spontaneous activity and ($\widetilde{\text{HH}}$) for the propagation failures, since these are the prominent phenomena in the respective dynamical system.

We aim to  propose a simple statistical estimator that allows for detection of both spontaneous activity and propagation failures. A first educated guess might suggest that checking for certain threshold crossings of the maximum height of the membrane potential, i.\,e. $\sup_{x \in (0,L)} u(x) > \theta$, is a good choice. Note that such a criterion has been used in \cite{Faisal} to detect arrival times of action potentials. However, we suggest a different method using the following linear functional of the (shifted) membrane potential,
\[
\Phi(u) \df \int_0^L u(x) - u^\ast \dx.
\]
This describes the area below the pulse of the membrane potential shifted by the resting potential $u^\ast$. Note that we can always change variables so that in the following we assume w.\,l.\,o.\,g. $u^\ast = 0$. We choose the estimator $\Phi$ over any other pointwise criterion as e.\,g. the supremum for the following reasons. First, $\Phi$ is a linear functional of only one observable. Second, the action potential is not a point charge that propagates along the axon but it is rather spread out along some part of it that may reach up to a few cm in length. Thus, a global criterion as imposed by $\Phi$ is more reasonable than a pointwise one. Moreover local fluctuations due to the noise should have a less pronounced effect. Third, $\Phi$ is not sensitive to fluctuations in the phase of the traveling pulse, which will be explained in the discussion section.

\begin{figure}[ht]
\centering
\includegraphics{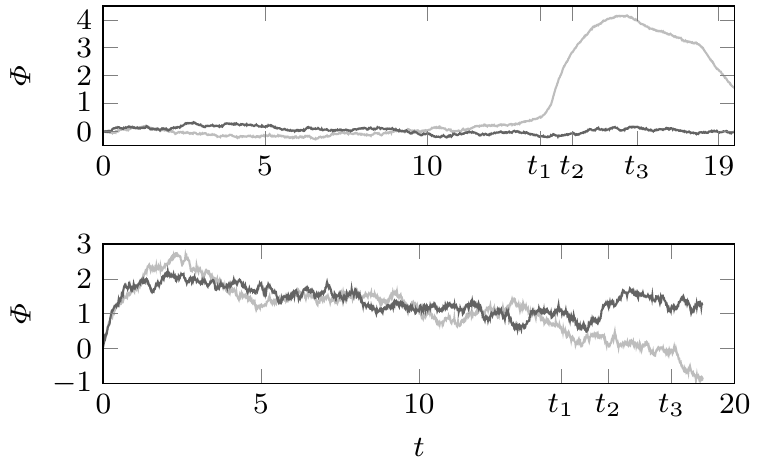}
\caption[]{The evolution of the area $\Phi$. (Top) For the same realizations as in Figure \ref{fig:sa_example} using (HH). Spontaneous activity (light gray) and fluctuations around the resting potential (dark gray). (Bottom) For the same realizations as in Figure \ref{fig:pf_example} using ($\widetilde{\text{HH}}$). Pulse with propagation failure (light gray), pulse without failure (dark gray).}
\label{fig:area}
\end{figure}

 Consider the deterministic solution (i.\,e. $\sigma=0$) $\hat{u}$ that is a traveling pulse and denote $\hat{\Phi} \df \Phi(\hat{u})$. As long as the pulse is formed, this quantity should stay more or less constant. In the following, with abuse of notation we use $\Phi(u) \df \Phi(u)/ \hat{\Phi}$. Concerning the example paths in Figure \ref{fig:sa_example} and \ref{fig:pf_example} we can look at the corresponding time evolution of the area $\Phi$, see Figure \ref{fig:area}.
 
\subsection{Spontaneous activity}

Since the estimator $\Phi$ reliably discriminates between no, one or more pulses, it can be used to observe the probability of emerging secondary pulses. In this scenario, starting the model (HH) at the resting potential without any input signal through the Neumann boundary condition, we observe the solution for the time $T$ the deterministic pulse $\hat{u}$ would need to reach the right boundary. For a given critical value $\theta$ we define the event $\sup_{t \in [0, T]} \Phi\big( u^\sigma(t)\big) \geq \theta$ as {\em spontaneous activity} for the noise amplitude $\sigma.$ Similar, the probability of spontaneous activity is
\[
s_\sigma \df \Prob{\sup_{t \in [0, T]} \Phi\big( u^\sigma(t)\big) \geq \theta}.
\]
In this definition, the threshold $\theta$ still has to be specified. Experience with different parameter sets and other neuron models have shown that a suitable threshold depends heavily on these. Suitable is used here in the sense that the estimator indeed detects an emerging action potential when there is one.

\begin{figure}[ht]
\centering
\includegraphics{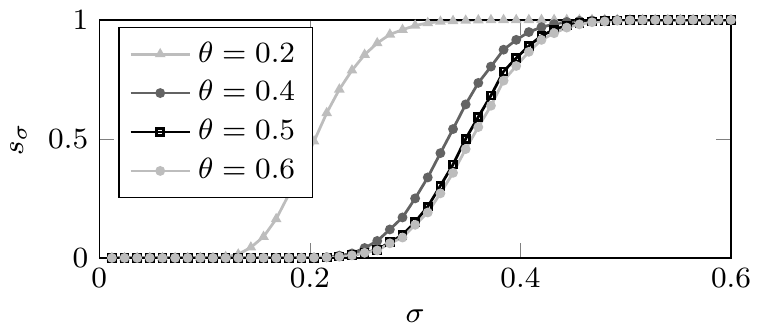}
\caption[]{Plot of $s_\sigma$ vs. $\sigma$ for different threshold values using the model (HH).}
\label{fig:sa}
\end{figure}

In the following we use $T=60$ and $M=10\,000$ realizations of (HH) to estimate $s_\sigma$. Figure \ref{fig:sa} shows that the curve $\sigma \mapsto s_\sigma$ shifts to the right as $\theta$ is increased and stays unchanged for $\theta\geq 0.52$, which is in this case the suitable threshold to detect spontaneous activity. 

\subsection{Propagation failure}

\begin{figure}[ht]
\centering
\includegraphics{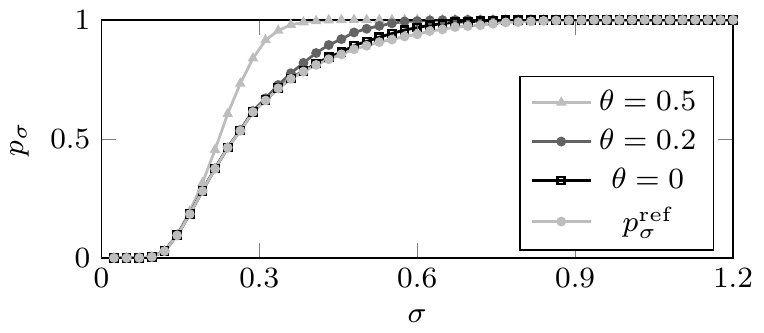}
\caption[]{Plot of $p_\sigma$ vs. $\sigma$ for different values of $\theta$ using the model ($\widetilde{\text{HH}}$).}
\label{fig:pf}
\end{figure}

Obviously, we can use $\Phi$ the other way round to detect a propagation failure using the model ($\widetilde{\text{HH}}$). Thus, we are in principle able to easily reproduce and generalize the observations made in \cite{TuckwellFHN} in terms of variation of parameters, models and the number of Monte-Carlo realizations.  Let $T_0 > 0$ be a given, fixed initialization time until the pulse is formed. Also, recall that $T$ denotes the time when the pulse has reached the boundary. Given a threshold $\theta$ we define the event $\sup_{t \in [T_0, T]} \Phi\big(u^\sigma(t)\big) - \hat{\Phi} > \theta$ as a \emph{propagation failure} for the noise amplitude $\sigma$. Similar, the probability of propagation failure is
\[
p_\sigma \df  \Prob{\sup_{t \in [T_0, T]} \Phi\big(u^\sigma(t)\big) - \hat{\Phi} > \theta}.
\]
\begin{remark}
Numerically the stopping time $T$ is implemented as follows. The axon is extended using a noiseless cable at the right boundary that allows to keep track of the pulse even if it already has left the original part of the axon. Applying the estimator $\Phi$ on both the noisy and noiseless part makes it possible to determine whether and when a pulse has successfully reached the axon terminal. With this, we can compute a reference value $p_\sigma^{\text{ref}}$ to evaluate the quality of the estimator $\Phi$.
\end{remark}
With $T_0 = 10$ Figure \ref{fig:pf} shows the probability of propagation failure $p_\sigma$ versus $\sigma$ for different threshold values compared to $p_\sigma^{\text{ref}}$. As $\theta$ decreases, the curves converge to the reference curve. In particular, $\theta = 0$ seems like a suitable threshold in this scenario.

\section{Model reduction}

Obtaining an analytical expression for $p_\sigma$ and $s_\sigma$ is out of reach, considering these are the exit time probabilities of a nonlinear infinite dimensional problem. 
However, one can use the linear stability assumptions of both pulse and resting state made in Section 2.3 to obtain a simplified model. In this part we show that a model reduction is indeed possible and propose a simple, one-dimensional equation that mimics the behavior of the original problem and is able to capture the desired quantities, such as the probabilities of propagation failure and spontaneous activity. This has the following implications: First, the computational costs are reduced and second, we obtain a simplified analytical expression in terms of classical, known quantities.

In view of assumption \eqref{eq:StabilityPulse} our arguments for the use of $\Phi$ can be strengthened by a simple observation. Let $\mathbbm{1}_u = (\mathbbm{1}, 0, \dots)^T$ be the constant function equal to $1$ in the $u$-component, then 
\[
\scp{\mathbbm{1}_u}{\dxt}_\mathcal{H} = \sfrac{\mathrm{d}}{\dt} \scp{\mathbbm{1}_u}{\hat{X}(t)}_\mathcal{H} = 0
\]
for $t \in [T_0, T]$ since the integral is invariant to translation of the pulse. Thus, $\mathbbm{1}_u \in \bot_t$ for all $t \in [T_0, T]$.

The implications of this are the following. Consider the solution $Z(t)$ to the linearization of \eqref{eq:AbstractSPDE} neglecting all higher order terms. In particular, $Z(t)$ is an Ornstein-Uhlenbeck process on $\mathcal{H}$. Writing $T(t,s) = \exp[\int_s^t A + \nabla F(\hat{X}(r)) \dr]$ for the exponential of the linear operator, the solution can be written using Duhamel's principle as
\[
Z(t) = T(t,0) Z(0) + \sigma \int_0^t T(t,s) \,\mathrm{d}\mathbb{W}(s).
\]
$Z(t)$ is a Gaussian process, uniquely characterized by its mean and variance
\begin{align*}
\EV{Z(t)} &= T(t,0) Z(0),\\
\Var{Z(t)} &= \sigma^2 \int_0^t T(t,s) T(t,s)^\ast \ds,
\end{align*}
where $\ast$ denotes the adjoint operator. Now, recall $\Phi(u) = \int_0^L u \dx$, hence $\Phi(u(t) - \hat{u}(t)) = \scp{u(t) - \hat{u}(t)}{\mathbbm{1}}_H \approx \scp{Z(t)}{\mathbbm{1}_u}_\mathcal{H}$. In particular, this is a linear functional of $Z(t)$. Since $Z$ is Gaussian, so is $\scp{Z(t)}{\mathbbm{1}_u}_\mathcal{H}$ with mean and variance
\begin{align*}
\EV{\scp{Z(t)}{\mathbbm{1}_u}_\mathcal{H}} &= \scp{T(t,0) Z(0)}{\mathbbm{1}_u}_\mathcal{H},\\
\Var{\scp{Z(t)}{\mathbbm{1}_u}_\mathcal{H}} &= \sigma^2 \int_0^t \scp{T(t,s)T(t,s)^\ast \mathbbm{1}_u}{\mathbbm{1}_u}_\mathcal{H} \ds.
\end{align*}
Now, recall \eqref{eq:StabilityPulse}, i.e. the linear stability assumption for the pulse state. Note that $\mathbbm{1}_u \in \bot_t$, i.e. orthogonal to the direction of pulse propagation, and therefore $\hat{C} \langle \mathbbm{1}_u , d_{\hat{X}}(t)\rangle^2_\mathcal{H} = 0$. Hence, the linear operator $T(t,s)$ satisfies the following inequality:
\[ \norm{T(t,s) \mathbbm{1}_u}_\mathcal{H} \leq \mathrm{e}^{-\hat{\kappa}(t-s)} \norm{\mathbbm{1}_u}_\mathcal{H}. \]
In particular it follows that
\begin{align*}
\EV{\scp{Z(t)}{\mathbbm{1}_u}_\mathcal{H}} &\leq \mathrm{e}^{-\hat{\kappa}t} \norm{Z(0)}_\mathcal{H} \norm{\mathbbm{1}_u}_\mathcal{H}\\
&\leq \sqrt{L} \mathrm{e}^{-\hat{\kappa}t} \norm{Z(0)}_\mathcal{H}.
\end{align*}
Of course, this implies $\EV{\scp{Z(t)}{\mathbbm{1}_u}_\mathcal{H}} \to 0$, which is one of the main advantages of choosing the estimator $\Phi$. In contrast to this, the squared $L^2$-norm $\norm{u(t) - \hat{u}(t)}_H^2$ or also $\sup_{x \in (0,L)} \abs{u(t,x) - \hat{u}(t,x)}$ might also serve as a measure of how close $u$ is to the pulse solution. However, both will not converge to $0$, since due to the noise $u$ will never be adapted to the right phase of $\hat{u}$. In our approach, we integrate the difference $u - \hat{u}$ with respect to a function orthogonal to the direction of propagation, hence our estimator does not perceive any phase shift and is locally exponentially stable around $0$. Concerning the variance, we compute
\begin{align*}
\Var{\scp{Z(t)}{\mathbbm{1}_u}_\mathcal{H}} &= \sigma^2 \int_0^t \norm{T(t,s) \mathbbm{1}_u}_\mathcal{H}^2 \ds\\
&\leq \sigma^2 \int_0^t \mathrm{e}^{-2\hat{\kappa} s} \ds \norm{\mathbbm{1}_u}_\mathcal{H}^2 \leq \frac{\sigma^2 L}{2 \hat{\kappa}}.
\end{align*}
With the considerations above, the following Ansatz for a scalar valued stochastic differential equation for $\Phi$ is reasonable.
\[
\mathrm{d} \Phi \big(u(t) - \hat{u}(t)\big) = - \alpha \Phi\big(u(t) - \hat{u}(t)\big) \dt + \tilde{\sigma}\,\mathrm{d}\beta(t),
\]
where $\beta(t) \df \sqrt{L}^{-1} \scp{W(t)}{\mathbbm{1}}_H$ defines a real-valued Brownian motion and $\tilde{\sigma} \df \sqrt{L} \sigma$. Using linearity of $\Phi$, $\hat{\Phi} \df \Phi(\hat{u}(t))$ and $\Phi(t) \df \Phi(u(t))$ it follows that
\begin{equation}\label{eq:OUapprox}
\mathrm{d}\Phi(t) = \alpha \big(\hat{\Phi} - \Phi(t)\big) \dt + \tilde{\sigma} \,\mathrm{d}\beta(t), \quad \Phi(0) = \hat{\Phi}
\end{equation}
is the approximating dynamics, a simple, one-dimensional Ornstein-Uhlenbeck process around the mean $\hat{\Phi}$. Also, $p_\sigma$ can be approximated by the exit time probability
\[
\tilde{p}_\sigma \df \Prob{\sup_{t \in [T_0, T_1]} \Phi^\sigma(t) - \hat{\Phi} > \theta},
\]
$T_1 = \EV{T}$, that is a first passage time of the Ornstein-Uhlenbeck process. These are intensively studied in relation to stochastic LIF neurons, see \cite{Alili, Sacerdote}, and are in addition easily accessible numerically.

In this Ansatz, the whole complexity of the SPDE dynamics is reduced to the parameter $\alpha$ and the solution to \eqref{eq:OUapprox} can be written down explicitly as
\[
\Phi(t) = \hat{\Phi} + \big( \Phi(0) - \hat{\Phi}\big) \mathrm{e}^{-\alpha t} + \tilde{\sigma} \int_0^t \mathrm{e}^{-\alpha ( t-s)} \,\mathrm{d}\beta(s).
\]
Assuming the validity of this linear approximation, which will be true for small $\sigma$, we can estimate $\alpha$ using mean and variance of $\Phi(t)$. In particular, 
\begin{align*}
\EV{\Phi(t)}  &= \hat{\Phi} + \big( \Phi(0) - \hat{\Phi}\big) \mathrm{e}^{-\alpha t},\\
\Var{\Phi(t)} &= \EV{ \Bigg( \tilde{\sigma} \int_0^t \mathrm{e}^{-\alpha (t-s)} \,\mathrm{d}\beta(s) \Bigg)^2}\\
&=\tilde{\sigma}^2 \int_0^t \mathrm{e}^{-2\alpha s} \ds = \frac{L \sigma^2}{2 \alpha} \Big( 1 - \mathrm{e}^{-2\alpha t}\Big).
\end{align*}
Hence, $\Var{\Phi(t)} \to \sfrac{L\sigma^2}{2\alpha}$ as $t \to \infty$ can be used to estimate $\alpha$ for large $t$, in our simulations $t = 45$, thus the difference to the limit is negligible. We apply the standard variance estimator
\[
\mathrm{Var}_M \df \tfrac{1}{M-1} \sum_{k=1}^M \big(\Phi_k(t) - \overline{\Phi}_M\big)^2, \quad \overline{\Phi}_M \df \tfrac1M \sum_{k=1}^M \Phi_k(t)
\]
for $\sigma = 0.024$, the smallest $\sigma$ used in the simulations before. We arrive at
\begin{equation}\label{eq:SpectralGap}
\alpha_M \df \frac{L \sigma^2}{20 \mathrm{Var}_M^\sigma} \approx 0.404,
\end{equation}
with again $M=10\,000$ realizations.

\begin{figure}[ht]
\centering
\includegraphics{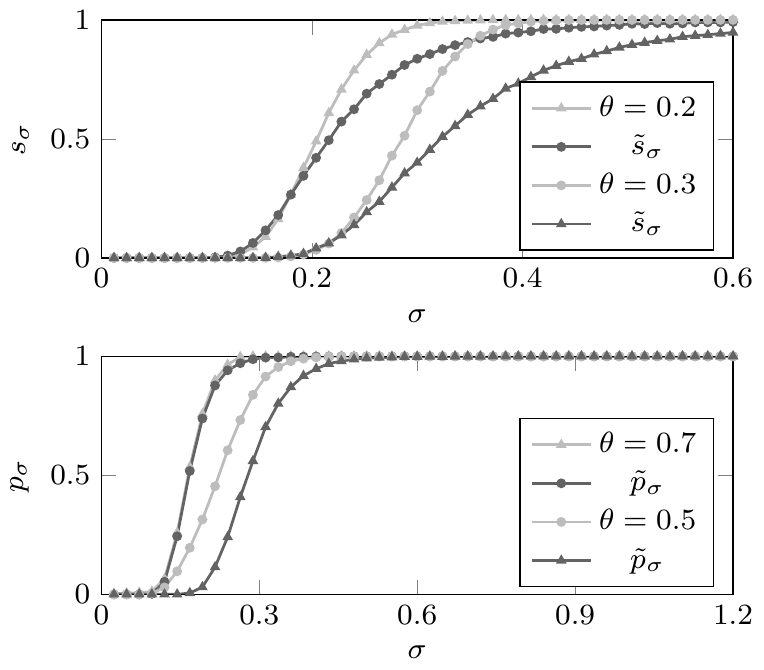}
\caption[]{Top plot: $\tilde{s}_\sigma$ vs. $\sigma$ for two threshold values $\theta$ in comparison to the corresponding values for $s_\sigma$ using the model (HH). Bottom plot: $\tilde{p}_\sigma$ vs. $\sigma$ for two threshold values $\theta$ in comparison to the corresponding values for $p_\sigma$ using ($\widetilde{\text{HH}}$).}
\label{fig:ou}
\end{figure}

Using the linearization around $X^\ast$ and the same Ansatz, we propose a similar Ornstein-Uhlenbeck process, whose hitting probabilities approximate $s_\sigma$. With $\Phi(t) \df \Phi(u(t)) = \scp{u(t)}{\mathbbm{1}}_H$ and, of course, $\Phi(u^\ast) = 0$ this reads as
\begin{equation}\label{eq:OUapprox2}
\mathrm{d}\Phi(t) = -\beta \Phi(t) \dt + \sigma \,\mathrm{d}\beta(t), \quad \Phi(0) = 0.
\end{equation}
Also, $\EV{\Phi(t)} = 0$ and $\Var{\Phi(t)} = \sfrac{L\sigma^2}{2\beta} (1 - \mathrm{e}^{-2\beta t})$ and we estimate the rate $\beta$ using $\sigma = 0.012$ via
\begin{equation}
\beta_M \df \frac{L\sigma^2}{20 \mathrm{Var}^\sigma_M} \approx 0.334
\end{equation}
with $M=10\,000$ realizations. Figure \ref{fig:ou} shows the probabilities $\tilde{p}_\sigma$ and 
\[
\tilde{s}_\sigma \df \Prob{\sup_{t \in [T_0, T_1]} \Phi^\sigma(t) > \theta}
\]
as a function of $\sigma$ for different thresholds $\theta$ compared to the probabilities obtained using the SPDE. Note that the approximation becomes worse as $\theta$ and $\sigma$ increase, which is expected since then the solution approaches the other equilibrium state and the linearization is not valid anymore.

\section{Discussion}
In this article, we have introduced a method to compute probabilities for spontaneous activity and propagation failure in a consistent way with underlying spatially extended, conductance-based neuronal models, based on certain statistical properties of the membrane potential. Since the action potential in the neuron's axon is not a point charge, but rather spread out in space, we advertise the use of a non-local criterion such as the one using $\Phi$. It may be interesting to find out how the axon's length and diameter influence the quality of detection, since these are the relevant parameters concerning the width of an action potential.

A further reduction in computational costs and a simplified analytical description can be achieved performing a model reduction with respect to the chosen estimator $\Phi$ in a consistent way with the underlying spatially extended neuronal model. This is based on its linearization at the resting potential (resp. the traveling action potential) and allows to approximate the probabilities for spontaneous activity and propagation failure in terms of (classical) hitting time probabilities of one-dimensional linear stochastic differential equations. Since the linearization is valid only locally, the approximations $\tilde{p}_\sigma$ and $\tilde{s}_\sigma$ become worse for growing $\theta$ and $\sigma$ as shown in Figure \ref{fig:ou}. For reasonable small $\theta$ and $\sigma$ however, the hitting probabilities of the one-dimensional stochastic differential equations are a solid approximation to the full nonlinear, infinite dimensional SPDE. On the other hand, Fig. 7 also shows that the model reduction can be used to find upper bounds for $s_\sigma$ resp. $p_\sigma$ over a considerably larger range of $\sigma$. 


In this study, we used the modified model ($\widetilde{\text{HH}}$) to illustrate propagation failures. Although \cite{Faisal} found action potential propagation to be very secure with less than $1\%$ failures, we have shown that little change in parameters produce a dynamical system with a totally different behavior. More precisely, rising slightly the sodium inactivation rate as in the modified Hodgkin-Huxley system ($\widetilde{HH}$) lowers excitability of the neuron and increases the probability of propagation failure. It may 
even become the predominant feature over spontaneous activation, similar to the case of the FitzHugh-Nagumo system, see \cite{TuckwellFHN}. It would be a interesting to see whether this computational fact could be confirmed in experiments.

As generalizations, we may incorporate more general noise, e.\,g. as suggested in \cite{GoldwynSheaBrown} for the Hodgkin-Huxley model, and study how this affects the signal transmission. Note, that in the development of this study we have used e.\,g. conductance noise as presented in \cite{Linaro}. This does not qualitatively change the behavior concerning $p_\sigma$ and $s_\sigma$, but should be analyzed in comparison to the results in \cite{Faisal} for the Hodgkin-Huxley equations with ion channels modeled via Markov chains. Future work will also be concerned with the effect of noise on the generation of repetitive spiking, see \cite{TuckwellJost}, and the estimation of the speed of propagation.

\end{document}